\documentclass[pra,aps,twocolumn,showpacs]{revtex4}
\usepackage{amsfonts,amsmath,bm}

\newcommand{\rl}{\rangle\!\langle}

\usepackage{graphicx}

\begin{document}

\title{Collective spontaneous emission from pairs of quantum dots:
long-range vs. short-range couplings} 
\author{Wildan Abdussalam}
\author{Pawe{\l} Machnikowski}
 \email{Pawel.Machnikowski@pwr.wroc.pl}  
\affiliation{Institute of Physics, Wroc{\l}aw University of
Technology, 50-370 Wroc{\l}aw, Poland}

\begin{abstract}
We study the spontaneous emission from a coherently delocalized exciton state
in a double quantum dot as a function of the distance between
the dots, focusing on the similarities and
differences between the cases of radiative (long-range, dipole) and tunnel
coupling between the excitons in the dots. We show that
there may be no qualitative difference between the collective emission
induced by these two coupling types in spite of their 
essentially different physical properties.
\end{abstract}


\maketitle

\section{Introduction} 
\label{sec:intro}

Excitons delocalized in closely spaced quantum dots (QDs) recombine
in a different way than in a single QD \cite{bardot05,borri03}.
This effect is at least partly due to collective 
interaction of the two emitters with the quantum radiation field
\cite{sitek07a}. For non-interacting dots this collective effect appears
only if the interband transition energies in the two QDs differ by no
more than the emission line width, which requires the dots to be
nearly identical, beyond the current technological
possibilities. However, coupling between the dots restores the
collective nature of the emission and leads
to accelerated or slowed down emission even for dots with different
transitions energies, which is manifested in the optical response from
these systems \cite{sitek09b,sitek09a}.  

The two major couplings that may appear in a system of QDs are due to
Coulomb interactions and carrier tunneling. The former results from the
coupling between the interband dipole moments associated with the
excitons in the dots (sometimes referred to as the \textit{dispersion
  force}) \cite{stephen64}. It has a long-range nature, with the typical
$1/R^{3}$ behavior at short distances (actually, this singularity is
removed for charges distributed in a finite volume
\cite{machnikowski09a}) and an oscillating tail with an envelope
decaying as $1/R$ at distances larger than the resonant wave
length. The tunnel coupling is a short-range interaction, which
vanishes exponentially at distances on the order of a few nanometers.

In this contribution, we study the spontaneous emission from an exciton
confined in a double quantum dot, focusing on the similarities and
differences between the cases of radiative (long-range, dipole) and tunnel
coupling between the excitons in the dots. We show that for strictly
identical dots the oscillating nature of the dipole coupling on long
distances leads to non-monotonic dependence of the radiative decay
rate on the inter-dot separation. However, for a double dot system
with a realistic, 
technologically feasible mismatch of transition energies, the
collective effects disappear completely well before these
oscillations become relevant. In both cases, there is no qualitative
difference between the emission induced by the long-range dipole
interaction and that due to short-range tunnel couplings with
appropriately chosen (but realistic) parameters, in spite of their
essentially different physical origin and properties.

The paper is organized as follows. In Sec.~\ref{sec:model}, we
describe the model of the system. Next, in
Sec.~\ref{sec:results},  we present and discuss the results of
numerical simulations. Finally, Sec.~\ref{sec:concl}
concludes the paper. 

\section{Model} 
\label{sec:model}

We consider two QDs placed in the $xy$ plane and shifted by a vector
$\bm{r}_{12}$. Each QD is modeled as a two-level system (empty 
dot and one exciton).
The Hilbert space of the double-dot system in our model is then spanned by
the empty dot state $|00\rangle$, the two single exciton states
$|10\rangle,|01\rangle$ corresponding to the exciton in the first and
second dot, respectively, and the ``molecular biexciton'' state $|11\rangle$.
The transition
energies for the interband transitions in the two dots are
\begin{displaymath}
E_{1}=\overline{E}+\epsilon,\quad
E_{2}=\overline{E}-\epsilon.
\end{displaymath}
The dots are coupled by an interaction $V$ which can be either
of dipole--dipole character (long-range \textit{dispersion force}) or
result from carrier tunneling (short-range, exponentially decaying
interaction). We introduce the transition (``exciton annihilation'')
operators for the two dots, $\sigma_{1}=|00\rl 10|+|01\rl 11|$, 
$\sigma_{2}=|00\rl 01|+|10\rl 11|$ and the exciton number operators 
$\hat{n}_{\alpha}=\sigma_{\alpha}^{\dag}\sigma_{\alpha}$,
$\alpha=1,2$. 
Using these operators, the Hamiltonian of the double-dot system is
written in the frame rotating with the frequency $\overline{E}/\hbar$
in the form  
\begin{displaymath}
H_{0}=\epsilon \left( \hat{n}_{1}-\hat{n}_{2} \right) 
+V\left( \sigma_{1}^{\dag}\sigma_{2}+\sigma_{2}^{\dag}\sigma_{1}
\right) +E_{\mathrm{B}} \hat{n}_{1}\hat{n}_{2},
\end{displaymath}
where last term represents the biexciton shift.

The long-range dipole coupling is described by
\begin{displaymath}
V=V_{\mathrm{lr}}=-\hbar\Gamma_{0}G(k_{0}r_{12}),
\end{displaymath}
where 
\begin{displaymath}
\Gamma_{0}=
\frac{|d_{0}|^{2}k_{0}^{3}}{3\pi\varepsilon_{0}\varepsilon_{\mathrm{r}}},
\end{displaymath}
is the spontaneous emission (radiative recombination) rate for a
single dot, 
$\varepsilon_{0}$ is the vacuum permittivity,
$\varepsilon_{\mathrm{r}}$ is the relative dielectric constant of the
semiconductor, and
\begin{displaymath}
k_{0}=\frac{n\overline{E}}{\hbar c},
\end{displaymath}
where $c$ is the speed of light and
$n=\sqrt{\varepsilon_{\mathrm{r}}}$ is the refractive index of the
semiconductor, 
and
\begin{eqnarray*}
G(x) & = & \frac{3}{4}\left[
-\left( 1-|\hat{\bm{d}}\cdot\hat{\bm{r}}_{12}|^{2}  \right)
\frac{\cos x}{x} \right. \\
&& \left.+\left( 1-3|\hat{\bm{d}}\cdot\hat{\bm{r}}_{12}|^{2}  \right)
\left( \frac{\sin x}{x^{2}}+\frac{\cos x}{x^{3}} \right)
 \right],
\end{eqnarray*}
where
$\hat{\bm{r}}_{12}=\bm{r}_{12}/r_{12}$ and
$\hat{\bm{d}}=\bm{d}/d$, where $\bm{d}$ is the interband matrix
element of the dipole moment operator which is assumed identical for
both dots. For a heavy hole exciton,
$\bm{d}=(d_{0}/\sqrt{2}) [1 , \pm i, 0]^{T}$, so that for a vector
$\bm{r}_{12}$ in the $xy$ plane one has
$|\hat{\bm{d}}\cdot\hat{\bm{r}}_{12}|^{2}=1/2$.
The tunnel coupling is described by 
\begin{displaymath}
V=V_{\mathrm{sr}}=V_{0}e^{-r_{12}/r_{0}}.
\end{displaymath}

The effect of the coupling to the radiation field is
accounted for by including the dissipative term in the evolution equations,
which describes radiative recombination of excitons.
The equation of evolution of the density matrix is then given
by \cite{lehmberg70a} 
\begin{equation}\label{evol}
\dot{\rho}=-\frac{i}{\hbar}[H_{0},\rho]+
\sum_{\alpha,\beta=1}^{2}\Gamma_{\alpha\beta}\left[ 
\sigma_{\alpha}\rho\sigma_{\beta}^{\dag}
-\frac{1}{2}\left\{ \sigma_{\alpha}^{\dag}\sigma_{\beta},\rho \right\}_{+}
 \right], 
\end{equation}
where $\Gamma_{11}=\Gamma_{22}=\Gamma_{0},\quad
\Gamma_{12}=\Gamma_{21}=\Gamma_{0}F(k_{0}r_{\alpha\beta})$,
with 
\begin{eqnarray*}
F(x) & = & \frac{3}{2}\left[
\left( 1-|\hat{\bm{d}}\cdot\hat{\bm{r}}_{12}|^{2}  \right)
\frac{\sin x}{x} \right.\\
&&\left.+\left( 1-3|\hat{\bm{d}}\cdot\hat{\bm{r}}_{12}|^{2}  \right)
\left( \frac{\cos x}{x^{2}}-\frac{\sin x}{x^{3}} \right)
 \right],
\end{eqnarray*}
and $\{\ldots,\ldots\}_{+}$ denotes the anticommutator. The diagonal
decay rates $\Gamma_{\alpha\alpha}$ describe the emission properties
from a single dot, while the off-diagonal terms
$\Gamma_{\alpha\beta}$, $\alpha\neq\beta$, account for the interference
of emission amplitudes resulting from the interaction with a common
reservoir and are responsible for the collective effects in the emission.

In our simulations, we  use the parameters for a typical InAs/GaAs QD system:
$\Gamma_{0}=1$~ns$^{-1}$, $n=3.3$, $\overline{E}=1.3$~eV. For the
tunnel coupling we choose the amplitude $V_{0}=2.19$~meV and the range
$r_{0}=2.03$~nm, which makes the values for the tunnel and dipole
couplings similar for inter-dot distances around 6~nm.

\begin{figure}[tb]
\begin{center}
\includegraphics[width=85mm]{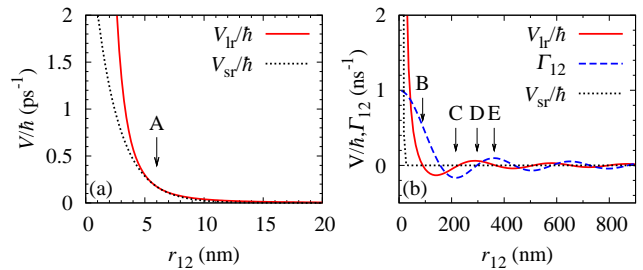}
\end{center}
\caption{\label{fig:GV} The interference term of the decay rate $\Gamma_{12}$ and
  the short- and long-range coupling amplitudes
  $V_{\mathrm{lr}},V_{\mathrm{sr}}$ as a function of the inter-dot
  distance. In (a), the small distance section is shown, while in (b)
  the oscillating tail at larger distances is visible. Note the
  different scales in (a) and (b).}
\end{figure}

The values of the two couplings as well as the interference term of
the decay rate  $\Gamma_{12}$ are plotted as a function of the
distance between the dots in Fig.~\ref{fig:GV}. In this figure we mark
the distance values for which the decay will be discussed in the next section.

\section{Results} 
\label{sec:results}

\begin{figure}[tb]
\begin{center}
\includegraphics[width=85mm]{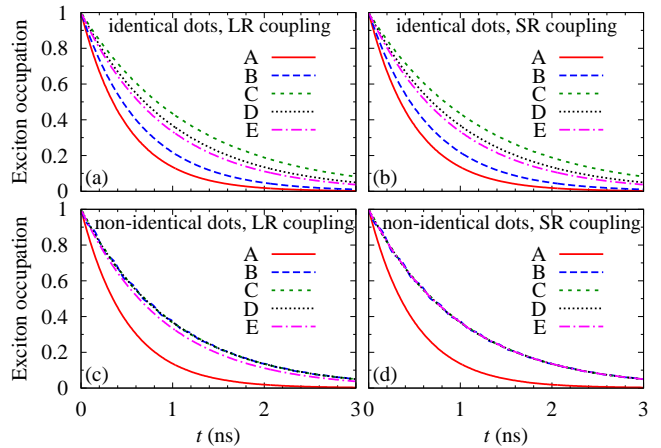}
\end{center}
\caption{\label{fig:decay}The exciton occupation (the average number of
  excitons in the system) for an initial single-exciton state corresponding to a
  coherently delocalized superposition. (a) and (b) show the evolution
for a pair of identical dots coupled by long-range dipole forces and
by short-range tunnel coupling, respectively. (c) and (d) show the
evolution for a pair of non-identical dots, for the two kinds of
couplings as previously.  The labels A,...,E refer to the
  values of the inter-dot distance marked in Fig.~\ref{fig:GV}.}
\end{figure}

In Fig.~\ref{fig:decay} we show the results of the numerical simulations based
on Eq.~\eqref{evol}. On each plot, the average number of excitons in
the system is shown as a function of time for identical dots
($\epsilon=0$) and for slightly non-identical dots with
$\epsilon=0.01$~meV. The initial state in all the cases is chosen to
be $(|01\rangle+|10\rangle)/\sqrt{2}$. We study the decay of exciton population for
various distances between the dots and compare the evolution for the
two kinds of couplings.

For identical dots, the exciton decay time for the delocalized initial
state strongly depends on the distance between the dots. This is due
to the oscillations and decay of the interference term
$\Gamma_{12}$. For dots placed at a short distance (case A), $\Gamma_{12}\sim
\Gamma_{0}$ and the decay has a strongly collective character, which
manifested by the faster emission visible in
Figs.~\ref{fig:decay}(a,b) \cite{sitek07a}. The collective
effect gets weaker as the distance between the dots grows and
$\Gamma_{12}$ decreases (B). For some values of the distance,
$\Gamma_{12}<0$ (C). Then, the amplitudes for photon emission from the
two dots interfere destructively and the decay gets slower than the usual
exponential decay with the rate $\Gamma_{0}$ (the initial state
becomes subradiant). Whenever $\Gamma_{12}=0$, the decay rate is the
same as for an individual dot (D). Comparison of
Figs.~\ref{fig:decay}(a) and (b) shows that for identical dots, these effects
do not depend on the coupling and are therefore the same, irrespective
of the presence and physical nature of the interaction between the
dots. 

For dots that differ by the relatively small transition energy
mismatch of $2\epsilon=0.02$~meV, almost all this non-monotonic dependence of
the emission rate on the distance disappears. The reason is that the
oscillations of the interference term take place in the distance range
where the coupling between the dots is very weak and is dominated already
by a small energy mismatch assumed here, which destroys collectivity of the
emission process \cite{sitek07a}. The only exception is the smallest
distance shown in 
this plot, where the coupling is sufficiently strong. By comparing
Figs.~\ref{fig:decay}(c) and \ref{fig:decay}(d) one can see that also
in this case, the evolution of the exciton occupation is nearly the
same for both systems. Here the tunnel coupling parameters have been
deliberately chosen to assure the same coupling strength around the
6~nm distance. At larger distances both couplings are negligible
compared to the energy mismatch.

\section{Conclusions}
\label{sec:concl}

We have shown that the radiative decay of exciton occupation in a pair
of coupled quantum dots depends on the distance between the dots as a
result of the spatial dependence of the interference term governing
the interaction with the quantum electromagnetic field. For
non-identical dots, the emission rate depends on the interplay of the
energy mismatch between the dots and the coupling between
them. Although the two kinds of couplings that are present in
the system (tunneling and dipole interaction) have essentially
different physical nature and properties, they may lead to the same
dynamics of the observed collective emission.

We believe that these findings may shed some light on the
interpretation of the experiment \cite{scheibner07} in which enhanced
emission was 
observed in a quantum dot ensemble in which the dipole coupling
energies on the typical inter-dot distances were much smaller than the
average transition energy mismatch between the dots. Indeed, as we
have shown, the
tunnel coupling leads to the same effect as the long-range dipole
interaction but it can be stronger than the latter at short
distances. Hence, it seems very 
likely that short-range tunnel coupling between some pairs of dots can
be responsible for the observed collective emission effect in QD ensembles.

\textbf{Acknowledgment:} This work was supported by the Foundation for
Polish Science under the TEAM programme, co-financed by the European Regional Development Fund.


\begin{thebibliography}{1}

\bibitem{bardot05}
C. Bardot, M. Schwab, M. Bayer, S. Fafard, Z. Wasilewski, and P. Hawrylak,
  Phys. Rev. B {\bf 72},  035314  (2005).

\bibitem{borri03}
P. Borri, W. Langbein, U. Woggon, M. Schwab, M. Bayer, S. Fafard, Z.
  Wasilewski, and P. Hawrylak, Phys. Rev. Lett. {\bf 91},  267401  (2003).

\bibitem{sitek07a}
A. Sitek and P. Machnikowski, Phys. Rev. B {\bf 75},  035328  (2007).

\bibitem{sitek09b}
A. Sitek and P. Machnikowski, Phys. Rev. B {\bf 80},  115319  (2009).

\bibitem{sitek09a}
A. Sitek and P. Machnikowski, Phys. Rev. B {\bf 80},  115301  (2009).

\bibitem{stephen64}
M.~J. Stephen, J. Chem. Phys. {\bf 40},  669  (1964).

\bibitem{machnikowski09a}
P. Machnikowski and E. Rozbcki, Phys. Stat. Sol. (b) {\bf 246},  320  (2009).

\bibitem{lehmberg70a}
R.~H. Lehmberg, Phys. Rev. A {\bf 2},  883  (1970).

\bibitem{scheibner07}
M. Scheibner, T. Schmidt, L. Worschech, A. Forchel, G. Bacher, T. Passow, and
  D. Hommel, Nature Physics {\bf 3},  106  (2007).

\end{thebibliography}

\end{document}